# Forty-Four Pass Fibre Optic Loop for Improving the Sensitivity of Surface Plasmon Resonance Sensors

**Chin B Su and Jun Kameoka**

Department of Electrical and Computer Engineering, Texas A&M University, College Station, TX 77843

E-mail : su@ece.tamu.edu

**Abstract**

A forty-four pass fibre optic surface plasmon resonance sensor that enhances detection sensitivity according to the number of passes is demonstrated for the first time. The technique employs a fibre optic recirculation loop that passes the detection spot forty-four times, thus enhancing sensitivity by a factor of forty-four. Presently, the total number of passes is limited by the onset of lasing action of the recirculation loop. This technique offers a significant sensitivity improvement for various types of plasmon resonance sensors that may be used in chemical and biomolecule detections.

## 1. Introduction

Surface plasmon resonance (SPR) sensors have been used for chemical and biological sensing for more than two decades [1-3]. Surface plasmons are charge density waves that propagate along the surface of metals, and are created when an optical wave at some incident resonance angle impinges a dielectric-metal-detection region structure [4,5]. On resonance, the interaction of surface plasmon and the optical wave reduces the structure reflectivity. Simultaneously, the magnitude of the localized and exponentially decaying evanescent field on the detection region exhibits a maximum value at the gold-detection region interface. Perturbations of the evanescent field by the residing molecules shift the resonant angle [6].

SPR can be implemented using grating coupled [4-7] or prism coupled [8] ( Kretschmann) configuration. Here, we used the Kretschmann's configuration for demonstrating the 44-pass sensitivity. The standard Kretschmann's configuration consists of a light source and a detector on opposite sides of a prism that allows for one reflection (one pass) of the optical beam off the gold layer. In a n-pass system, the light impinges the detection region n times by multiple reflections. Recently, we reported on the performance of a SPR setup using a corner cube retroreflector replacing the photodetector to achieve four passes instead of one [9], and suggested an external fibre re-circulating loop method to increase the number of passes beyond four. In this paper, we demonstrate a working re-circulating loop that achieves forty-four passes, and we show the sensitivity scales with the number of passes. This technique detects both charged and uncharged analytes, in contrast to the field-assist technique reported in reference 10 that improves sensitivity by attracting only charged analytes to the detection region.

A brief analysis shows the enhanced sensitivity for more passes: If the detected power P ~ $R^n$, where R is the SPR reflectivity, then the fractional change in power, $\delta P/P$, due to a resonant angle shift, $\delta\theta$, is given by

$$\frac{\delta P}{P} = n\left(\frac{d\ln(R)}{d\theta}\right) \cdot \delta\theta \qquad (1)$$

which increases with the number of passes, n.

**2. Experimental Setup**

Figure 1 shows the SPR setup with the all-fibre recirculating loop. The SPR setup comprises a fibre optic collimator on one side of the prism and a mirror reflector on the other side that reflects the beam back into the fibre collimator, resulting in a stand-alone 2-pass configuration that has a lower optical back-coupling loss than the 4-pass configuration of reference 9. The lower loss eases the burden on the optical amplifier used in this experiment.

The principle of the forty-four pass operation is described below. Properties of fibre components used here are reviewed in reference 11. Pulse generator 1 drives a diode laser (LD) to produce an optical pulse train with about 5 % duty cycle. The pulse width is about 0.13 µS. Pulse generator 2 is gated by this pulse-train to produce a synchronize pulse-train with a much longer pulse width, T, as shown, the purpose of which will be described later. The optical pulse incident on fibre coupler FC1 is splitted into two pulses. One pulse propagates towards port 1 of the optical circulator after traversing a polarization controller, PC3, and a fibre delay line. The optical pulse-train proceeds towards the SPR setup by exiting port 2 of the optical circulator. The fibre collimator collimates the laser beam that impinges on the gold-coated BK-7 substrate. The beam reflected off the gold-coated substrate is reflected back to the fibre collimator by the mirror, retracing the original optical path. Thus, the SPR setup itself is a two-pass device. The optical pulse that is significantly reduced in amplitude due to SPR resonance effect and the back-coupling loss at the collimator re-enter the fibre loop via port 3 of the optical circulator. The pulse is amplified and restored to the initial amplitude by the erbium-doped fibre amplifier after passing through the electro-optic modulator (EOM). The pulse eventually reaches FC1 to complete one round-trip. This returned pulse is again splitted into two pulses by FC1, one pulse is detected by a detector/amplifier module (PD), and the other pulse travels towards the SPR to repeat the process. Therefore, the detector/amplifier module detects a series of periodic pulses. The first pulse does not pass the SPR device. The second pulse passes the SPR twice, and the third pulse passes the SPR four times etc. The period is determined by the round-trip time through the fibre loop and the SPR setup. The added fibre delay line in the fibre loop is simply to assure that the 0.13 µS re-circulating pulses are well separated in time.

The EOM functions as a loss-modulating optical switch. The switch is closed (low loss) when the gated electrical pulse applied to the RF port of the EOM is on, otherwise the switch is opened (high loss). The time duration of the gated pulse, T, determines the total number of passes of the SPR system. This switching action is important because the fiber loop with the optical amplifier comprises a fibre laser that can lase without any input, thus, destroying the function of the SPR. The periodic opening of the EOM switch prevents lasing from occurring, but, as a compromise, limits the maximum number of achievable passes. Appropriate adjustment of the three polarization controllers, PC1, PC2, PC3 ensures that the same optical polarization is maintained for every round-trip of the re-circulating pulse, and, also, the polarization is p-polarized at the SPR for exciting the surface plasmon [4].

**Experimental Results**

The basic SPR function is first verified by measuring its one-pass characteristics by disconnecting the recirculation loop and by temporarily replacing the reflecting mirror on the SPR setup by a photodetector. The reflectivity versus incident angle profile is shown in figure 2 when DI water is dispensed onto the gold surface. One obtained the familiar SPR curve with a minimum reflectivity at resonance occurring at an incident angle of about 62.6 (corresponds to 0° in figure 2). Our diode laser wavelength is 1.53 μm, compatible with erbium-doped fibre amplifier technology. At 1.53 μm wavelength, the resonance profile is sharper and the reflectivity dip is shallower than the response at traditionally used wavelength of 0.78 μm, which can be verified by simulation [9].

For multipass applications we rotate the collimator and the mirror to set the bias point at 0.17° (±0.02) below resonance, as indicated by the arrow in figure 2. The forty-four pass response is first performed with DI water (18 MΩ-cm quality) by dispensing it on the gold surface. The gain of the fiber amplifier is adjusted such that amplitudes of pulses for the case of water are approximately equal as shown by the lower trace in figure 3. Then the water is replaced with a 0.01% gram-salt/cc (1.7 mM) salt solution, and its pulse-train is measured. Both results are superimposed in figure 3. The presence of salt increases the solution's index, and according to SPR theory, shifts the resonant angle to slightly larger angle, causing an increase in the reflectivity and optical signal when the set bias angle is below the resonance angle. The total number of pulses is 22 and the corresponding number of passes is 44, as each round-trip of the pulse through the loop impinges the gold surface twice. Figure 3 reveals the differential increase of the pulse-amplitude with number of passes for salt solution over DI water. This result demonstrates the higher sensitivity for more passes.

It is noted that the increase in magnitude of the base-line with time, as observed in figure 3, is due to the temporal increase in amplified spontaneous emission from the fiber amplifier that ultimately will lead to self- lasing within the loop if the gated pulse width, T, is too long.

From figure 3, one deduces the fractional signal increase (dictated by equation 1) of salt solution with respect to DI water for the various pulses as , $(P_{s,m} - P_{w,m})/P_{w,m}$ , where $P_{s,m}$ and $P_{w,m}$ are measured peak amplitudes of the $m^{th}$ pulse for the salt solution and DI water respectively. The fractional increase, $(P_{s,m} - P_{w,m})/P_{w,m}$ , is plotted in figure 4 as solid circles. Note that for 2 passes, the fractional signal increase is too small to be determined. But for 20 passes the fractional signal increase is about 0.55 , and for 40 passes, the fractional signal increase is 1.2 , which is a factor of about 2 increase from 20 to 40 passes .

An SPR program is used for predicting the fractional increase for the various passes. Quantities that need to be calculated are $[(R_s)^{2m} - (R_w)^{2m}]/(R_w)^{2m}$, where $R_s$ and $R_w$ are one-pass reflectivities for salt solution and water respectively. The $m^{th}$ pulse (not counting the first one) passes the SPR 2m times, and the effective reflectivity is raised to 2m power. The program for calculating the reflectivity uses the transmission matrix method for plane waves. The original method can be found in reference 12. Basically, the method assumes an incident and reflected electromagnetic field in any individual layer of a specified multilayer dielectric (including metallic) structure. By matching tangential components of E and H fields at layers' boundaries, one obtains an angle-dependent reflectivity. Reflectivity, $R_w$ and $R_s$, are calculated for the "two structures": BK7- Ti -Au -water and BK7 -Ti – Au- salt solution. Water index $n_w$ is 1.3159, which is smaller at 1.5 μm than at visible wavelengths [14]. The salt solution index is taken as $n_w$ + δn, where δn is varied to fit to the data in figure 4. δn = 2.7 x$10^{-5}$ gives the best fit to the measured data. The predicted δn is obtained from the formula, $n_w$+ δn= $n_s$x +(1-x)$n_w$ [14] , where $n_s$ = 1.544 is the index of salt at 1.5μm, $n_w$ = 1.3159 is the index of water, and x = $10^{-4}$ (salt weight fraction) , the concentration of the salt solution. The predicted δn is 2.3 x$10^{-5}$, which agrees fairly well with the fitted value of δn = 2.7 x$10^{-5}$.

Other common parameters used in the calculation are : BK7 prism index : 1.50065. Ti complex index : 4.0 + j*3.8 (thickness 10 nm). Au complex index : 0.2 + j*10.2 (thickness 50 nm). These parameters at the 1.5 μm wavelength are obtained from reference 13.

**Conclusions**

In conclusion, a 44-pass all-fiber-optic technique for surface plasmon resonance sensor that enhances detection sensitivity according to the number of passes is introduced for the first time. The technique employs a fibre optic recirculation loop that passes the detection spot 44 times thus enhancing sensitivity by a factor of 44. The key to the successful implementation of the concept is the gated switch that turn off the fiber loop to suppress lasing effects. This technique offers significant sensitivity improvements over traditionally used one-pass plasmon resonance sensors.

Presently, the total number of passes is limited by the onset of lasing action occurring after time T (T in figure 3). An obvious method to significantly increase the number of passes beyond what has been achieve here is to shorten the optical pulse to accommodate more pulses within the T ≈10 µS time duration before amplified spontaneous emission becomes too serious. A corresponding increase in the detection bandwidth is also needed.

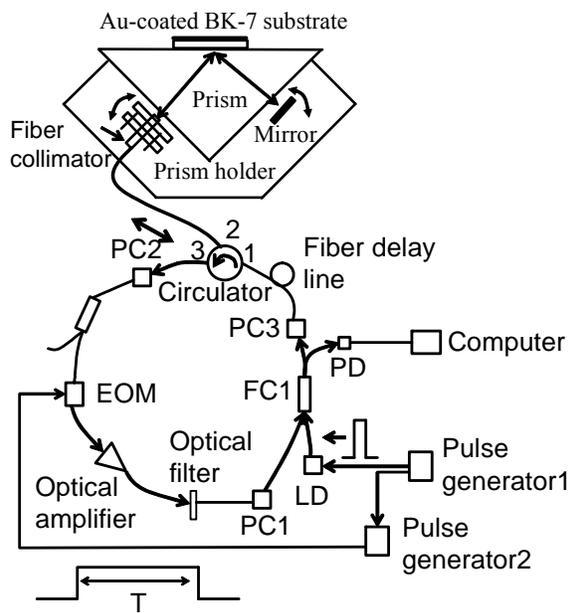

**Figure 1**

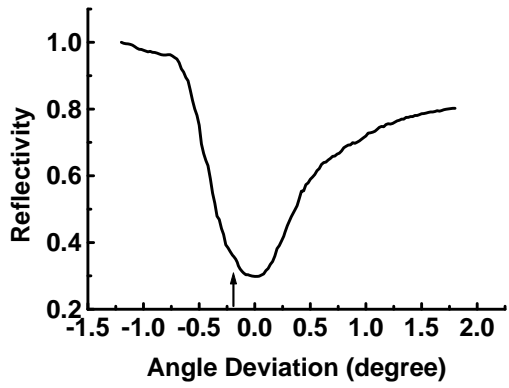

**Figure 2**

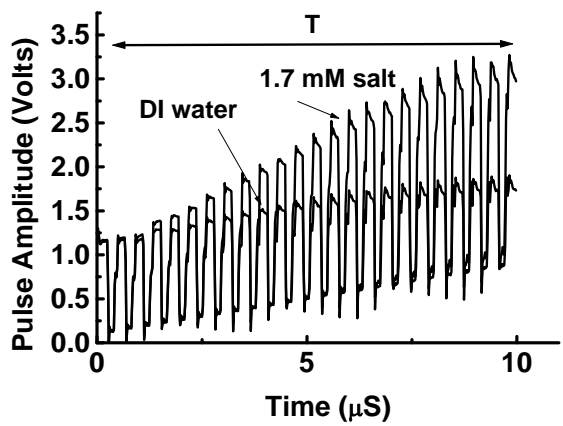

Figure 3

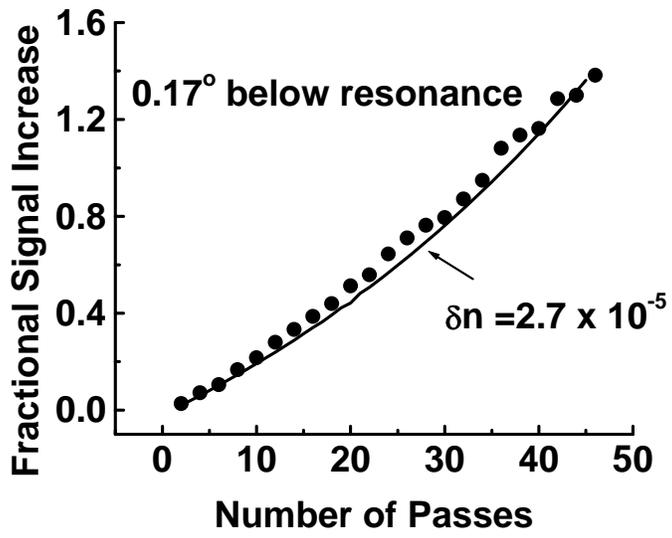

**Figure 4**

**Figure Captions**

Figure 1: The multipass setup. Top: SPR. Bottom: fiber loop. LD: laser diode. PD: photodiode. EOM: electrooptic modulator.

Figure 2: The one-pass reflectivity versus angle shift. Zero degree corresponds to the resonance angle at 62.6 °. The arrow represents the set bias angle at 0.17° below the resonant angle.

Figure 3 : Pulse amplitude versus number of passes for DI water (lower trace) and 1.7 mM salt solution (upper trace). T is the width of the gated pulse applied on the EOM

Figure 4: Fractional increase of signal for salt solution versus the number of passes. The points are fractional signal increase taken from figure 3. The line is the predicted increase as described in the text.